\documentclass[%
 reprint,
 amsmath,amssymb,
 aps,
pra,
]{revtex4-2}

\usepackage{graphicx}
\usepackage{dcolumn}
\usepackage{bm}
\usepackage{nicefrac}
\usepackage{amsmath}

\begin{document}

\preprint{APS/123-QED}

\title{Three-Dimensional and Selective Displacement Sensing of a Levitated Nanoparticle via Spatial Mode Decomposition}

\author{Thomas J. Dinter,$^{1}$ Reece Roberts,$^{1, 2}$ Thomas Volz,$^{1, 2}$ Mikolaj K. Schmidt,$^{1, 2}$ Cyril Laplane,$^{1, 2, 3, \dag, }$}

\email{cyril.laplane@sydney.edu.au} 

\affiliation{$^1$School of Mathematical and Physical Sciences, Macquarie University, NSW 2109, Australia\\
$^2$ARC Centre for Engineered Quantum Systems (EQUS), Macquarie University, NSW 2109, Australia\\
$^3$Sydney Quantum Academy, Sydney, NSW 2006, Australia\\
$^{\dag}$Present address: School of Physics, The University of Sydney, NSW 2006, Australia}

\begin{abstract}
We propose and experimentally demonstrate a novel detection method that significantly improves the precision of real-time measurement of the three-dimensional displacement of a levitated dipolar scatterer. Our technique relies on spatial mode sorting of the light scattered by the levitated object, allowing us to selectively extract the position information of all translational degrees of freedom with minimal losses. To this end, we collect all the light back-scattered from a levitated nanoparticle using a parabolic mirror and couple it into a spatial mode sorter. We measure displacement sensitivities ($\sqrt{S_{\mathrm{imp}, x}}, \sqrt{S_{\mathrm{imp}, y}}, \sqrt{S_{\mathrm{imp}, z}}$) $=$ (1.7, 2.4, 1.0) $\times$ $10^{-14}$ m/$\sqrt{\mathrm{Hz}}$ below the zero-point motion ($x_{\mathrm{zpm}}, y_{\mathrm{zpm}}, z_{\mathrm{zpm}}$) $=$ (2.2, 2.4, 1.6) $\times$ $10^{-12}$ m of the levitated particle considered here. In the regime where environmental decoherence is not limited by gas collision we estimate that our method can reach measurement efficiencies of $(\eta_{^{\mathrm{tot}}}^{_{x}}, \eta_{^{\mathrm{tot}}}^{_{y}}, \eta_{^{\mathrm{tot}}}^{_{z}}) = (0.13, 0.18, 0.33) > \nicefrac{1}{9}$, which would enable the 3D motional quantum ground state of a levitated optomechanical system.
\end{abstract}

\maketitle

\section{Introduction}
Levitated optomechanical systems have recently entered the quantum regime, with several demonstrations having now achieved cooling to the center-of-mass (COM) motional quantum ground state \cite{delic_cooling_2020, magrini_real-time_2021, tebbenjohanns_quantum_2021, kamba_optical_2022}. Such platforms present an exciting avenue for designing tests of fundamental physics \cite{moore_search_2014, bose_spin_2017, afek_coherent_2022, kilian_dark_2024}, and in particular for matter-wave interferometry experiments with mesoscopic objects \cite{scala_matter-wave_2013} several orders of magnitude larger than the current state-of-the-art~\cite{fein_quantum_2019, arndt_testing_2014}. A common prerequisite for most of these proposals is to prepare the levitated mechanical oscillator in its motional quantum ground state. To achieve ground-state cooling, research efforts can be separated into two distinct approaches: passive cooling using coherent scattering in a cavity \cite{delic_cooling_2020, piotrowski_simultaneous_2023, ranfagni_two-dimensional_2022}, and active cooling with measurement-based control techniques \cite{magrini_real-time_2021, tebbenjohanns_quantum_2021}. In particular, the latter actively controls an object's mechanical motion by applying a position or velocity-dependent feedback force. Hence, measurement-based control techniques are highly dependent on the efficiency with which we can extract information on the system's dynamics.

It was recently understood that the COM motion of a levitated nanoparticle generates coupling between different spatial modes of the scattered light, such that information on the different motional DOFs is encoded in specific spatial radiation patterns, so-called \textsl{information radiation patterns} \cite{tebbenjohanns_optimal_2019, maurer_quantum_2023}. Because of this, it is difficult to construct a free-space detection system that can efficiently measure each motional DOF of a levitated nanoparticle simultaneously. Structured-light detection techniques have recently been used to improve the measurement efficiency of the transverse motion (perpendicular to the laser propagation axis) in optical tweezers systems \cite{madsen_ultrafast_2021, Li:23}. Fibre-based confocal detection coupled with conventional interferometric techniques has only been able to reach quantum-enabling measurement efficiencies for the longitudinal motion (parallel to the propagation axis of the optical tweezer) by employing an appropriately shaped local oscillator \cite{magrini_real-time_2021, tebbenjohanns_quantum_2021, kamba_optical_2022}. 
Here, we generalise and extend these techniques, noting that due to the anisotropic coupling between the mechanical DOFs and the scattered light, the task of reading out the nanoparticle's motion becomes akin to spatial mode sorting. This suggests that the toolbox of space-division demultiplexing (SPADE) presents an interesting avenue for performing quantum-limited imaging of an oscillator's mechanical modes.

We present a novel approach to displacement sensing that relies on decomposing the light scattered from a levitated nanoparticle in the appropriate basis of spatial modes. As a convenient demonstration of our method, we choose the linearly polarised (LP) mode basis of a commercially available spatial mode sorter \cite{noauthor_lpmux_nodate}, encoding information on each translational DOF in the amplitude modulations of a unique channel. The experimental setup relies on simple, telecommunication grade and commercially available components, providing a path towards a compact and fully integrated levitated optomechanical platform. We perform parametric feedback cooling~\cite{gieseler_subkelvin_2012} to temperatures of ${(T_{x}, T_{y}, T_{z}) = (74.9, 260.3, 88.8)\pm(3.1, 4.5, 4.7)}$~mK at a pressure of $2.8\times 10^{-5}$~mbar and measure displacement sensitivities ($\sqrt{S_{\mathrm{imp}, x}}, \sqrt{S_{\mathrm{imp}, y}}, \sqrt{S_{\mathrm{imp}, z}}) = (17.5, 23.8, 10.0)\pm(0.9, 1.3, 0.5)$~fm~/~$\sqrt{\mathrm{Hz}}$. We show that this technique allows us to reach the measurement efficiencies necessary for realizing the 3D motional quantum ground state of a levitated nanoparticle.

\begin{figure*}[ht]
    \centering
        \includegraphics[width=\linewidth]{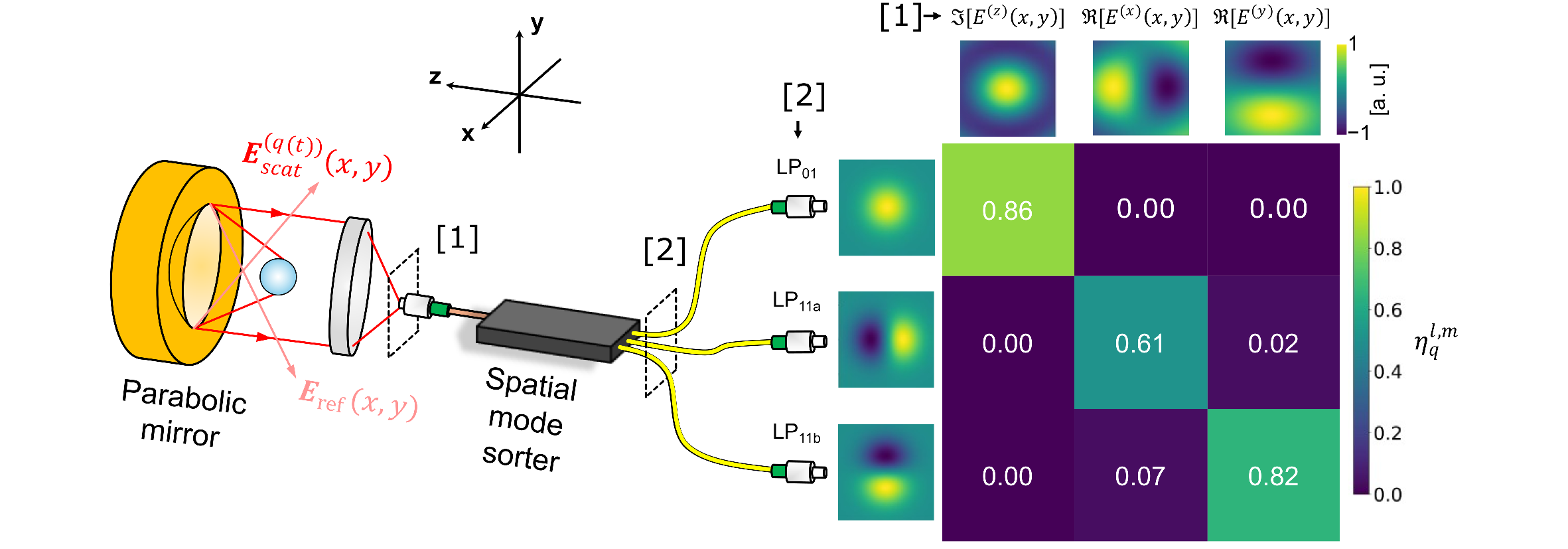}
        \caption{The operating principle of our displacement sensing technique using a spatial mode sorter. In the focal plane of the collection lens, labelled [1], the scattered field and reference field are coupled into the few-mode optical fiber input of the device. In the panel [1] on the right, we show the dominant components of the inelastically scattered field $E^{(q)}(x, y)$. Each component carries information on the nanoparticle's motion along the $\boldsymbol{\hat{e}}_{q}$ axis, and couples predominantly to a single LP mode. In the output plane, labelled [2], information on each translational degree of freedom $q(t)$ is stored in a unique channel of the sorter (see panel [2] for the plots of the electric fields of the three fiber modes). We find that the inelastically scattered field components, $\boldsymbol{E}^{(q)}(x, y)$, selectively couple to specific LP modes. Note that the LP$_{11\mathrm{a}/\mathrm{b}}$ modes are effectively de-coupled from the elastically scattered and reference fields, $\boldsymbol{E}_{0}(x, y)$ and $\boldsymbol{E}_{\mathrm{ref}}(x, y)$, respectively.
        }
        \label{FigOne}
\end{figure*}

\section{Sorting the Scattered Light}

We consider the case of a sub-wavelength nanoparticle, which is well-approximated as an isotropic dipolar scatterer, confined to the focus of a linearly ($x$) polarised trapping beam. To minimise polarisation mixing of the electric field at the focus while maintaining high collection efficiency of the parabolic mirror, we adjust the collimated beam diameter to a filling factor of $\approx$ 1. We model the geometry shown in Fig.~\ref{FigOne}, where the trapping beam is focused by a high numerical aperture (NA) parabolic mirror. This implementation presents a key advantage over traditional setups, in that it collects all of the back-scattered light, which is collimated and subsequently coupled into a few-mode optical fiber in the focal plane of the collimation/collection lens~\cite{vovrosh_parametric_2017}. At the output of the single-mode fibers, we measure the recovered light intensity to extract information on the nanoparticle's real-time position.

In the limit of small displacements, we can write the scattered field in the focal plane of the collection lens as a sum of elastically and inelastically scattered light (see Supplementary Material, Appendix~1). In particular, we find that only the inelastically scattered light is modulated by the nanoparticle's real time position and thus is the only one to contribute meaningfully to the measurement result. To first order, we approximate the total scattered field as
\begin{align}\label{E_scat}
    \boldsymbol{E}_{\mathrm{scat}}(x, y~; \boldsymbol{q}(t)) = \overbrace{\boldsymbol{E}_{0}(x, y)}^{\hidewidth \text{elastic}} + \sum_{\hidewidth q =x, y, z}\frac{q(t)}{\lambda_{0}}\overbrace{\boldsymbol{E}^{(q)}(x, y)}^{\hidewidth \text{inelastic}} ~ ,
\end{align}
where $\lambda_{0}$ refers to the trapping beam free-space wavelength, and $q(t)$ refers to the nanoparticle's position along the $\boldsymbol{\hat{e}}_{q}$ axis. In addition to these terms, we must also consider the unscattered and diverging part of the trapping beam, $\boldsymbol{E}_{\mathrm{ref}}(x, y)$. In the focal plane of the collection lens, this field is well-approximated by a plane wave and is typically tuned using an iris in front of the mirror such that ${|\boldsymbol{E}_{\mathrm{ref}}(x, y)| \gtrsim |\boldsymbol{E}_{\mathrm{scat}}(x, y~; \boldsymbol{q}(t))|}$.

\begin{figure*}[ht]
    \centering
        \includegraphics[width=\linewidth]{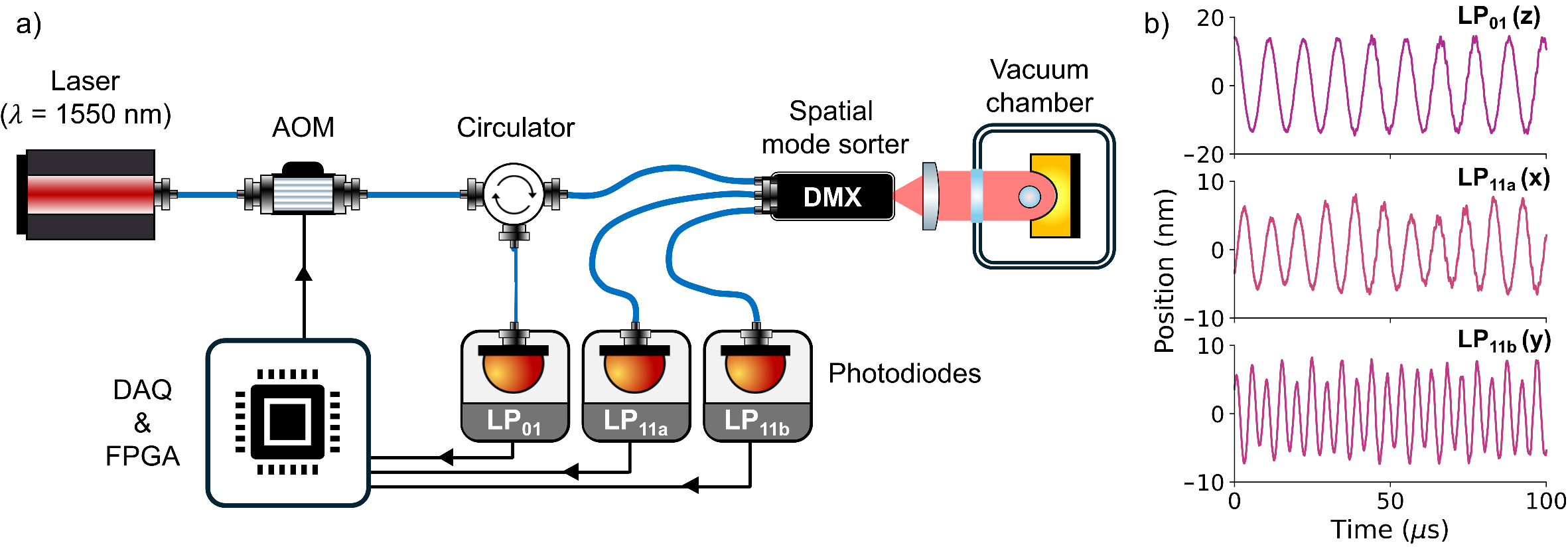}
        \caption{
        \textbf{(a)} Experimental setup. The output of a $\lambda = 1550$~nm laser is controlled by an acousto-optic modulator (AOM) and coupled into the LP$_{01}$ channel of a space-division demultiplexing photonic chip, producing optical tweezing at the focus of a high numerical aperture parabolic mirror. The back-scattered light from the levitated nanoparticle is then coupled back into the photonic chip. \textbf{(b)} Example of calibrated photodiode signals at the LP$_{01}$, LP$_{11\mathrm{a}}$ and LP$_{11\mathrm{b}}$ output channels of the spatial mode sorter at a pressure of 1.8 $\times$ 10$^{-3}$ mbar.
        }
        \label{FigTwo}
\end{figure*}

In Fig.~\ref{FigOne} we compare the dominant polarisation component of the inelastically scattered field $\boldsymbol{E}^{(q)}(x, y)$ to the lowest-order LP modes. We find that displacements parallel to the trapping axis ($\boldsymbol{\hat{e}}_{z}$) produce an electric field profile reminiscent of the LP$_{01}$ mode supported in optical fibers. Conversely, the fields produced for displacements along $\boldsymbol{\hat{e}}_{x}$ and $\boldsymbol{\hat{e}}_{y}$ are reminiscent of the LP$_{11\mathrm{a}}$ and LP$_{11\mathrm{b}}$ modes, respectively. Inspired by this observation, we choose to decompose the scattered field in the LP mode basis. This is achieved by coupling the light scattered from the nanoparticle into a commercially available space-division demultiplexing photonic chip \cite{noauthor_lpmux_nodate}, which functions as a spatial mode sorter. As depicted schematically in Fig.~\ref{FigOne}, this allows us to read out each translational DOF from amplitude modulations of a unique LP mode. Before quantifying this effect, however, let us first consider the experimental setup used to implement this selective measurement protocol.

\section{Experimental Setup}

We optically levitate a spherical SiO$_{2}$ nanoparticle of $\approx$ $130$~nm radius (mass $\approx$ 19 fg), using a simple experimental setup (illustrated in Fig.\ref{FigTwo}a). The trapping beam is tuned to a wavelength of $\lambda_{0}=1550$~nm with a power of $160$~mW. Upon loading the optical trap, the system is decompressed in a two-stage pumping process to a vacuum.

In the focal plane of the collimation/collection lens, the scattered field is coupled into the spatial mode sorter~\cite{noauthor_lpmux_nodate}. The input is a few-mode, graded-index optical fiber with a core diameter of approximately 8\,\textmu m, supporting the LP$_{01}$, LP$_{11\mathrm{a}}$ and LP$_{11\mathrm{b}}$ modes. Once excited, the electric field associated with each of these modes is transferred to a unique single-mode output fiber using a tapered mode-selective coupler \cite{riesen_ultra-broadband_2013, gross_three-dimensional_2014}. Each output channel is then sent to a photodiode (Thorlabs PDB450C), with a typical set of detected signals shown in Fig.~\ref{FigTwo}b. For an $x$-polarized trapping beam, we observe oscillation frequencies ${(\Omega_{z}, \Omega_{x}, \Omega_{y})~/~2\pi \approx (91, 111, 207)}$~kHz for each translational DOF. We attribute the other spectral features (at various linear combinations of $\Omega_{x}$, $\Omega_{y}$, and $\Omega_{z}$) to nonlinear mode-coupling between the different DOFs because of the trap anharmonicity. These features disappear when we cool the nanoparticle's motion (see Fig.\ref{FigFour}), which is further evidence of the thermally driven nonlinearities~\cite{gieseler_thermal_2013}.
For parametric feedback cooling, the signal of each photodiode is sent to an FPGA board (Red Pitaya), which functions as a digital phase-locked loop \cite{Jain2016}, with the recovered signal filtered and phase-shifted to generate a feedback signal at twice the natural oscillation frequency $2\Omega_{q}$. These feedback signals are used to modulate the power of the trapping beam via an acousto-optic modulator (AOM). 

\section{Results}

\begin{figure*}[ht]
    \centering
        \includegraphics[width=\linewidth]{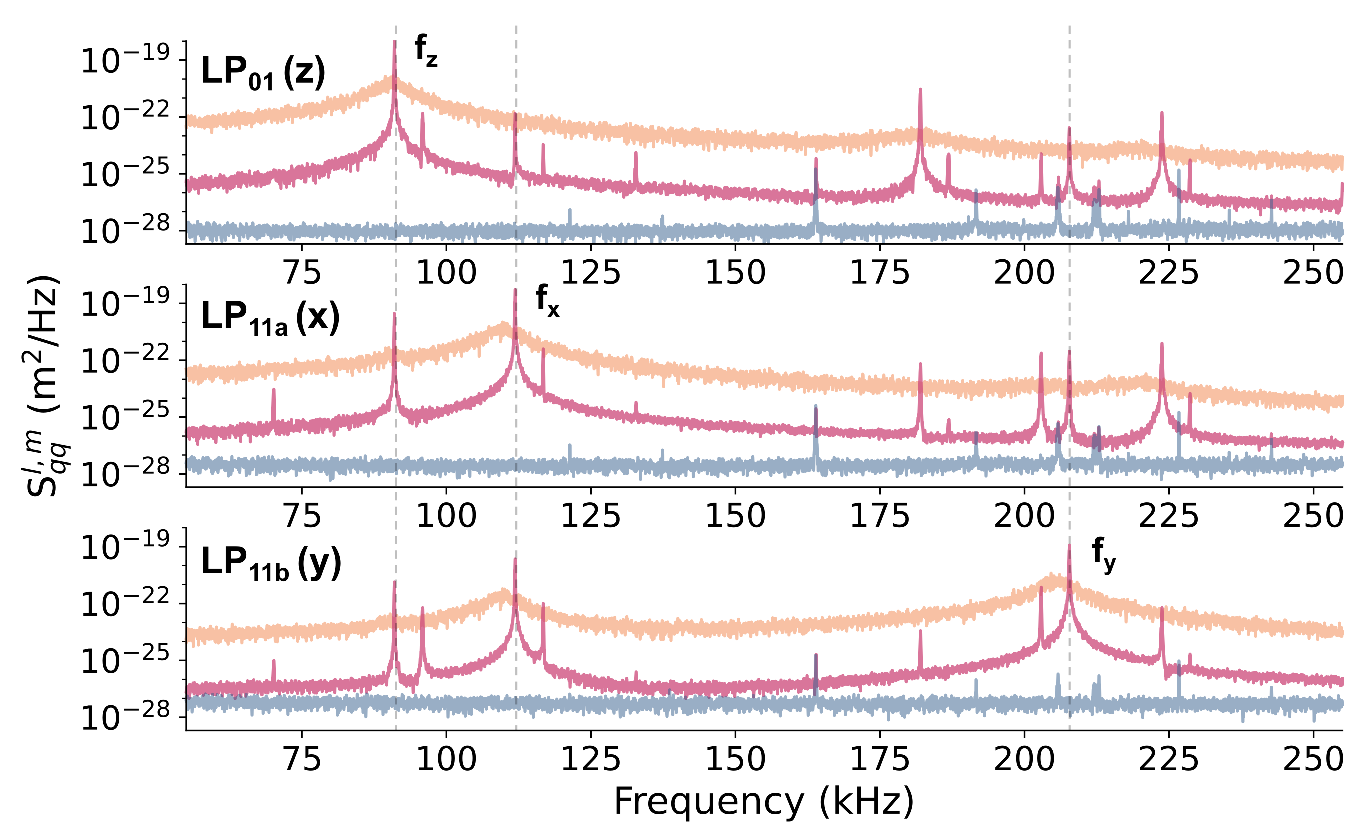}
        \caption{
        \textbf{(a)} Calibrated power spectral densities (PSDs) at the LP$_{01}$, LP$_{11\mathrm{a}}$ and LP$_{11\mathrm{b}}$ output channels of the spatial mode sorter. Note that each channel has contributions from primarily a translational degree of freedom, as predicted from the profile of $\boldsymbol{E}^{(q)}(x, y)$. PSDs are measured at a pressure of 11.3 mbar (peach) and 1.8 $\times$ 10$^{-3}$ mbar (ruby). The shot noise floor, measured when the trap is empty, is also shown (grey).
        }
        \label{FigThree}
\end{figure*}

\begin{figure*}[ht]
    \centering
        \includegraphics[width=\linewidth]{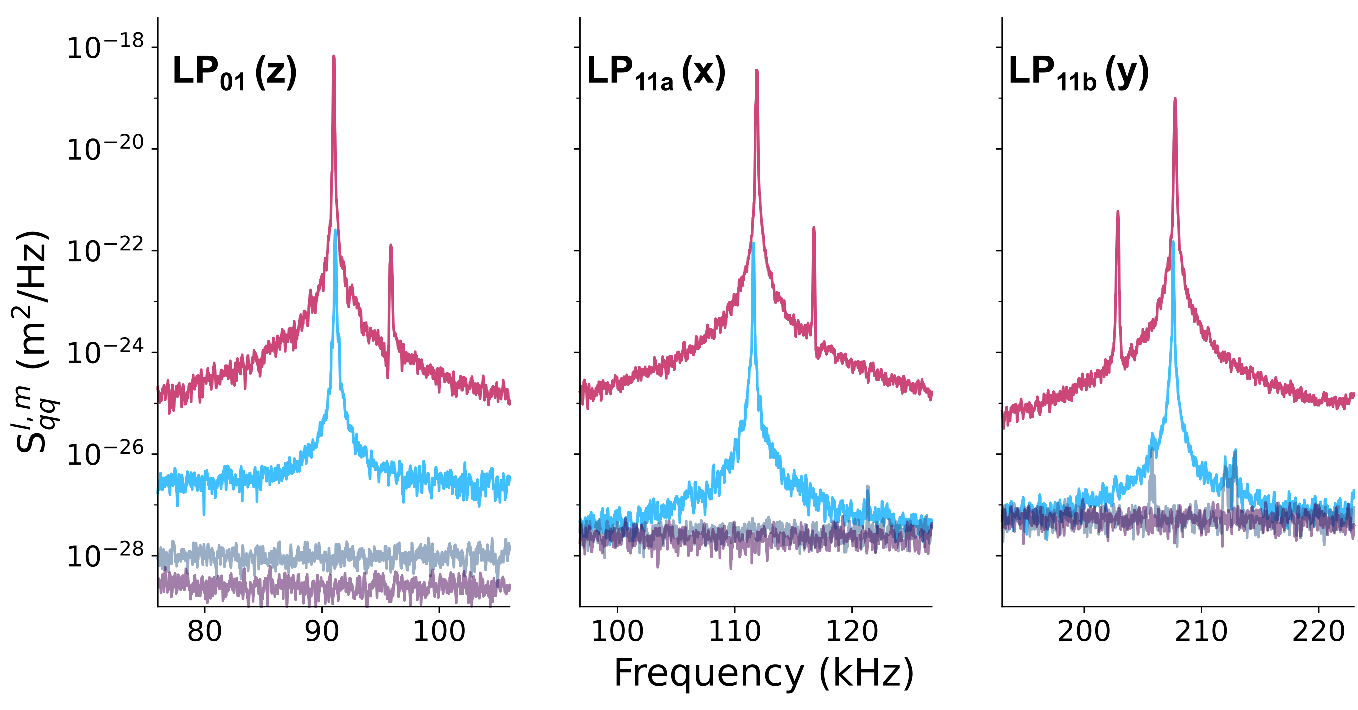}
        \caption{
        Measured PSDs without cooling at 1.8 $\times$ 10$^{-3}$ mbar (coral) and with parametric feedback cooling (blue) at a pressure of 2.8 $\times$ 10$^{-5}$ mbar showing ${(T_{z}, T_{x}, T_{y}) = (88.8, 74.9, 260.3)\pm(3.1, 4.5, 4.7)}$~mK. The shot noise floor, measured when the trap is empty, is shown (grey) as well as the detector noise (purple).
        }
        \label{FigFour}
\end{figure*}

In Fig.~\ref{FigThree} we plot the power spectral densities (PSDs) of the signals measured at each output channel of the spatial mode sorter. We find that each output channel contains information on primarily a single translational DOF, equivalent to the nanoparticle's motion along a particular spatial axis. This information is encoded in the PSDs labeled $S_{qq}(\Omega)$ for $q = x,y$ and $z$. We observe signal-to-noise ratio $\approx$ 80 dB (or above) for all three DOFs, already hinting at the high measurement efficiency of our system. In each channel, we observe extinction ratios ranging from 10 dB (for the LP11b channel) to 20 dB (for the LP01 channel), when comparing the dominant spectral feature with the second highest peak. In Fig.~\ref{FigThree} and Fig.\ref{FigFour}, we also show the noise floor measured in the absence of a trapped nanoparticle. Importantly, these extinction ratios match the technical specification supplied for the spatial mode sorter for the LP$_{11\mathrm{a}}$ and LP$_{11\mathrm{b}}$ channels, suggesting that we may be limited by the performance capabilities of the device. Slight misalignment between the axis of the demultiplexer (i.e. the orientation axis for the LP11 modes) and the main axis of the dipolar scatterer can also degrade the demultiplexing performance.

To quantify, and better understand the extent and limitations of this spatial mode sorting technique, we have numerically modelled the coupling efficiency of the scattered fields to each LP mode, defined as the spatial mode overlap integral \cite{wagner_coupling_1982, vamivakas_phase-sensitive_2007}:
\begin{align}\label{CouplingEq}
    \eta_{q}^{\ell,m} &= \frac{\left|\iint \big[\boldsymbol{E}^{(q)}(x, y)\big]^{*} \cdot \textbf{LP}_{\ell,m}(x, y) ~ \mathrm{d}x\mathrm{d}y\right|^2}{\iint \big|\boldsymbol{E}^{(q)}(x, y)\big|^{2}~\mathrm{d}x\mathrm{d}y \iint \big|\textbf{LP}_{\ell,m}(x,y)\big|^{2}~\mathrm{d}x\mathrm{d}y} , 
\end{align}
where $\textbf{LP}_{\ell,m}(x, y)$ refers to the electric field of the LP$_{\ell,m}$ mode, and integration is performed over the end-face of the input optical fiber. This procedure, including the corrections made to account for the geometry of our setup, can be found in the Supplementary Material, Appendix~2. Finally, we can write down analogous expressions for the coupling of the elastically scattered $\boldsymbol{E}_{\mathrm{0}}(x, y)$ and the reference $\boldsymbol{E}_{\mathrm{ref}}(x, y)$ fields also. Interestingly, we find that both fields are effectively decoupled from the LP$_{11\mathrm{a,b}}$ channels with coupling efficiencies $<$ 10$^{-6}$. The coefficients calculated for the particular experimental setup used here are reported in Fig.~\ref{FigOne}, with a detailed discussion on how we expect these quantities to relate to the measured PSDs found in the Supplementary Material, Appendix~4.

The predictions made in Fig.~\ref{FigOne} qualitatively match our experimental results, with the LP$_{11\mathrm{a}}$ (LP$_{11\mathrm{b}}$) mode coupling predominantly to the inelastically scattered field generated by displacements along $\boldsymbol{\hat{e}}_{x}$ ($\boldsymbol{\hat{e}}_{y}$), and the LP$_{01}$ mode containing only weak contributions from both the transverse DOFs. The elastically scattered and reference fields are only weakly coupled to the LP$_{11\mathrm{a,b}}$ modes, due to their spatial asymmetry. This would be disadvantageous in interferometric detection schemes, which rely on the reference field (that act as a local oscillator) to extract information from the phase of the scattered light. Here however, the information on the position of the nanoparticle for the transverse mode is encoded in the amplitude of the electric field coupled to the LP$_{11\mathrm{a,b}}$ channels. As shown in the Supplementary Material, Appendix~3, we find that this leads to a much simpler form for the measurement imprecision:
\begin{align}
    S^{\ell, m}_{\mathrm{imp}, q} &= \frac{\hbar k c}{8\pi}\sqrt{\frac{\mu_{0}}{\varepsilon_{0}}} \frac{1}{|\alpha_{q}|^{2}}\frac{\lambda_{0}^{2}}{\eta^{\ell,m}_{q}} ~ ,
\end{align}
where $\alpha_{q}$ is the complex amplitude of the inelastically scattered electric field components. Importantly, this result is independent of the overlap of the scattered and reference fields. This alleviates many of the technical challenges associated with conventional interferometric detection, namely the need to engineer a local oscillator (LO) which interferes optimally with the scattered light, and suggests that this technique enables greatly improved measurement efficiencies of the transverse DOFs. As an illustrative example, we note that the consequences of this weak coupling to the elastically scattered and LO fields can be observed clearly in Fig.~(\ref{FigFour}), where the noise floor of the PSDs in these channels closely matches the imprecision noise floor (empty trap). We emphasize here that the imprecision noise is still slightly above the detector noise. Finally, it is worth acknowledging that while the measurement efficiency of the nanoparticle's longitudinal motion does not gain any advantage from this technique, it is comparable to previous demonstrations of ground state cooling.

\section{Discussion}

We perform parametric feedback cooling to reduce the temperature of the COM motion of a levitated nanoparticle in Fig.~\ref{FigFour} and in particular to reduce the thermomechanical noise. The temperature $T_{q}$ achieved along the $\hat{\boldsymbol{e}}_{q}$ axis is determined by fitting the PSD measured at the appropriate output channel of the photonic chip (see Supplementary Material, Appendix~6). Using this setup, we cool the motion of the nanoparticle from thermal equilibrium ($300$~K) to a minimum temperature of ${(T_{z}, T_{x}, T_{y}) = (88.8, 74.9, 260.3)\pm(3.1, 4.5, 4.7)}$~mK, respectively at a pressure of 2.8 $\times$ 10$^{-5}$ mbar. Most importantly and particularly for the LP$_{11\mathrm{a,b}}$ channels, we find that we are limited by the shot noise of our experimental realisation with signal-to-noise ratio (SNR) that are still above 30 dB. In channel LP$_{01}$, we attribute the elevated noise floor of the PSD to the uncontrolled phase mismatch between the reference field (LO) and scattered field. This can be solved by using a phase-controlled external reference field as in conventional homodyne detection. Further our current experimental implementation is limited by the intensity noise of the laser and by employing a balanced detection we should be able to reduce the imprecision noise floor even further \cite{rashid_wigner_2017}.

We opted to calculate the measurement efficiencies of this setup for each translational DOF. To understand why this quantity is of interest, note that optical measurements necessarily disturb the system's natural evolution due to back-action. However, efficient measurements maximise the information gained per disturbance, and so can generate real-time feedback which not only controls the system's mechanical motion but also cancels the effects of said back-action. If we assume that there is no environmental losses (interaction with gas molecules) we can calculate the measurement efficiencies of our realisation from the measured noise floor shown in Fig.\ref{FigThree} and \ref{FigFour} (see Appendix 7 in the Supplentary Material) and we find ${(\eta_{^{\mathrm{tot}}}^{_{x}}, \eta_{^{\mathrm{tot}}}^{_{y}}, \eta_{^{\mathrm{tot}}}^{_{z}})_{\mathrm{exp}} = (0.10, 0.06, 0.31)}$. We outline an estimation of the various losses of our system in Table S1 of the Supplementary Material, Appendix 5 and we estimate the total measurement efficiencies for the translational DOFs, ${(\eta_{^{\mathrm{tot}}}^{_{x}}, \eta_{^{\mathrm{tot}}}^{_{y}}, \eta_{^{\mathrm{tot}}}^{_{z}})_{\mathrm{th}} = (0.13, 0.18, 0.33)} > \nicefrac{1}{9}$ in good agreement with our experimentally measured value. We emphasize here that this estimation is based on our current experimental implementation, and that small technical improvements such as using high quantum efficiency balanced photodetectors would enable even higher measurement efficiencies. Importantly, we report record measurement efficiencies for the transverse DOFs ($q=x, y$), not at the expense of the longitudinal DOF, for which we achieve a value comparable to previous experimental demonstrations of 1D ground state cooling~\cite{tebbenjohanns_quantum_2021, magrini_real-time_2021}. From these values, we can estimate the minimum achievable phonon occupancy for the nanoparticle's COM motion along each spatial axis~\cite{wiseman_quantum_2009, rossi_measurement-based_2018, tebbenjohanns_quantum_2021} in the limit of low pressures, where environmental information loss from random collisions with background gas molecules is negligible (${\approx 1\times10^{-8}}$~mbar)~\cite{clerk_introduction_2010}. Under these conditions, with an optimised feedback circuit, we expect to reach phonon occupancies of ${\bar{n}_{\mathrm{min}}^{q}=(1~/~\sqrt{\eta^{_{q}}_{^{\mathrm{tot}}}} - 1)~/~2}$. For the setup reported here, this is equivalent to ${(\bar{n}_{\mathrm{min}}^{x}, \bar{n}_{\mathrm{min}}^{y},\bar{n}_{\mathrm{min}}^{z}) = (0.89, 0.68, 0.37)}$ phonons. Hence, we believe that this is the first demonstration of a detection system capable of realizing the 3D motional quantum ground state of a levitated nanoparticle using measurement-based control techniques.

\section{Conclusion}

In summary, we have demonstrated a novel approach to measuring the center-of-mass motion of a levitated dipolar scatterer, wherein the scattered field couples into a spatial mode sorter. By decomposing the inelastically scattered contributions of this field in the LP mode basis of an optical fiber, we can isolate contributions from each translational DOF in the amplitude of orthogonal optical modes. Thanks to this technique, we report record high, quantum-enabling measurement efficiencies for the translational degrees of freedom, which suggests that our technique can realize the 3D motional quantum ground state.
Importantly, we believe that our technique can be extended to higher-order modes in the LP basis and generalized to other families such as Hermite-Gaussian (HG) modes, orbital angular momentum (OAM) modes, and Laguerre-Gaussian (LG) modes. This opens an exciting avenue for tracking the entire 6D (rotational and translational) motion of levitated objects with improved precision. Alternatively, one could harness the toolbox of structured light to match the inelastically scattered field of the levitated object perfectly, for example, by using a spatial light modulator. \\
It is worth acknowledging that although the present demonstration focuses on the field of levitated optomechanics, the technique has the potential to be employed with other trapped microscopic objects such as trapped ions or neutral atoms where independently and directly accessing their different motional states, i.e. bosonic modes, could prove practical for hybrid qubit-oscillator quantum information processing \cite{fluhmann_encoding_2019, navickas_experimental_2024, liu_hybrid_2024}.\\
Finally, the ability of the spatial mode sorter to isolate the scattered electric field associated with particular motional degrees of freedom is an essential step towards the passive, all-optical coherent feedback control of a levitated particle in free space. Close to the ground state, coherent feedback has the potential to improve cooling by avoiding measurement and feedback noise \cite{ernzer_optical_2023} and provides an avenue to generate mechanical squeezing \cite{karg_remote_2019}.
\textit{Note: }The authors have become aware of a related work where a similar technique was used to achieve quantum-limited read out of a nanomechanical resonator's torsional mode~\cite{choi_quantum_limited_2024}.

\section*{Acknowledgements}
Authors acknowledge fruitful discussions with Thiago Guerreiro, Michael Steel and Massimiliano Rossi. This work was supported by the Australian Research Council Centre of Excellence for Engineered Quantum Systems (Grant No. CE170100009) and Lockheed Martin. C. L. acknowledges support from the Sydney Quantum Academy Postdoctoral Research Fellowship and the Office of National Intelligence (ID250100445). M. K. S. acknowledges support from the Australian Research Council Discovery Early Career Researcher Award (DE220101272).

\section*{Data availability}
The data that support the findings of this study are available from the corresponding author upon reasonable request. 

\appendix

\section{Decomposition of the scattered field}\label{sec:FieldDerivation}

\begin{figure*}[ht]
    \centering
        \includegraphics[width=0.7\linewidth]{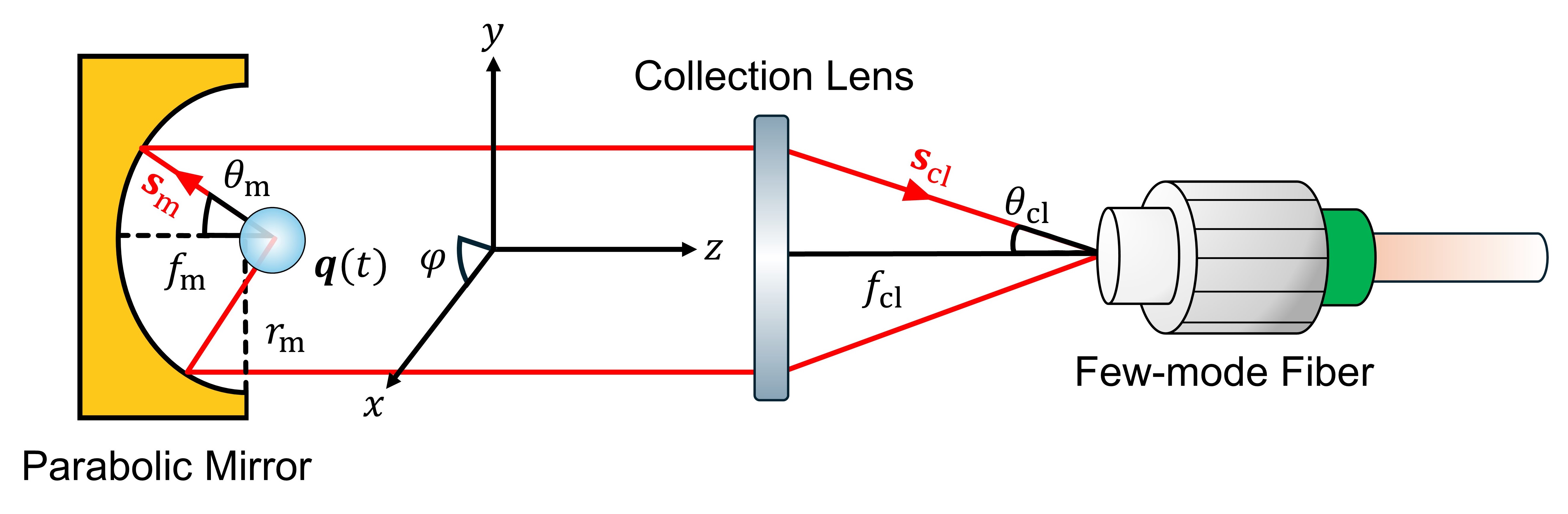}
        \caption{
        Geometry of the parabolic mirror being modelled, with key features labelled. A nanoparticle trapped at the diffraction-limited focal point of the parabolic mirror scatters a portion of the incident light, which is then coupled into the few-mode fiber input of a space-division demultiplexing photonic chip.
        }
        \label{ParabolicMirrorFig}
\end{figure*}

\noindent In this Appendix we provide the explicit expressions for the field scattered from a levitated dipolar scatterer, having been transformed by the parabolic mirror and collection lens composing the optical trap. In the Fraunhofer approximation, the scattered field is analogous to the far-field of an electric dipole, with an appropriate phase-shift:
\begin{align}\label{RawScatteredField}
    \boldsymbol{E}_{\mathrm{scat}}(\omega, \boldsymbol{r} ~;~ \boldsymbol{q}(t)) \approx \boldsymbol{E}_{\mathrm{dip}}(\omega, \boldsymbol{r})~\mathrm{exp}\left[-ik\left(\frac{\boldsymbol{r}\cdot\boldsymbol{q}(t)}{|\boldsymbol{r}|}\right)\right] ~ ,
\end{align}
where
\begin{align}
    \boldsymbol{E}_{\mathrm{dip}}(\omega, \boldsymbol{r}) &= \frac{k^{2}}{4\pi\varepsilon_{0}} \frac{e^{ik|\boldsymbol{r}|}}{|\boldsymbol{r}|}\big[\boldsymbol{r}\times\boldsymbol{p}(\omega, \boldsymbol{r})\big]\times\boldsymbol{r} ~ .
\end{align}
Here, $\boldsymbol{q}(t)$ refers to the nanoparticle's real-time displacement from the focal point of the parabolic mirror, and $\boldsymbol{p}(\omega, \boldsymbol{r})$ to the induced dipole moment, which we assume is aligned parallel to the polarization axis of the trapping beam. Moving forward, we drop the frequency dependence of these terms, and choose to focus on the case of a monochromatic trapping beam. We note that Eq.~(\ref{RawScatteredField}) does not describe the electric field observed at the input to the space-division demultiplexing photonic chip, however. Instead, the scattered field is transformed by a series of optical elements, resulting in modifications to both the spatially-varying phase and amplitude. In particular, for the optical trap design considered here, the back-scattered component of the field is first collimated by a gold-coated, high-numerical aperture (NA) parabolic mirror. We write the effective NA of this device as being of the form
\begin{align}
    \mathrm{NA} &= 1 - \mathrm{cos}\left(\mathrm{arctan}\left[\frac{r_{\mathrm{m}}}{f_{\mathrm{m}} - \frac{r_{\mathrm{m}}^{2}}{4f_{\mathrm{m}}}}\right]\right) ~ , 
\end{align}
where $f_{\mathrm{m}}$ and $r_{\mathrm{m}}$ are the focal length (1~mm) and radius (2~mm) of the parabolic mirror, respectively~\cite{vovrosh_parametric_2017}. The collimated field then propagates to some collection lens of focal length $f_{\mathrm{cl}}$ (25~mm) and is subsequently focused in the input plane of the photonic chip. A schematic of this setup is shown in Fig.~\ref{ParabolicMirrorFig} below.

This problem, of imaging the light scattered from a dipolar source, has been studied extensively in the context of confocal microscopy~\cite{lieb_high_2001}, where it is typically addressed using angular spectrum theory. For the setup considered here, we find a time-independent component of the electric field, in the focal plane of the collection lens, of the form
\begin{widetext}
\begin{align}
    \boldsymbol{E}_{\mathrm{scat}}(x, y~; \boldsymbol{q}(t)) &= -\frac{ik^{3}f_{\mathrm{cl}}}{16\pi^{3}\varepsilon_{0}} \int^{\theta_{\mathrm{cl}}^{\mathrm{max}}}_{0} \int^{2\pi}_{0} \frac{[1+\mathrm{cos}\left(\theta_{\mathrm{m}}\right)]}{2f_{\mathrm{m}}} \sqrt{\mathrm{cos}\left(\theta_{\mathrm{cl}}\right)} \left[(\boldsymbol{p}\cdot\boldsymbol{e}_{\mathrm{m}}^{\parallel})\cdot\boldsymbol{e}_{\mathrm{cl}}^{\parallel} - (\boldsymbol{p}\cdot\boldsymbol{e}_{\perp})\cdot\boldsymbol{e}_{\perp}\right] \nonumber  \\ 
    &\hspace{35mm} \times\mathrm{exp}\Big[-ik(x~\mathrm{sin}(\theta_{\mathrm{cl}})\mathrm{cos}(\varphi) + y~\mathrm{sin}(\theta_{\mathrm{cl}})\mathrm{sin}(\varphi) + (\hat{\boldsymbol{e}}_{q}\cdot\boldsymbol{s}_{\mathrm{m}})q(t))\Big] \mathrm{sin}(\theta_{\mathrm{cl}})~\mathrm{d}\theta_{\mathrm{cl}}\mathrm{d}\varphi ~ , \label{FieldExpression}
\end{align}
\end{widetext}
where we have introduced the relevant unit vectors, 
\begin{align}
    \boldsymbol{e}_{\perp} &= \Big[-\mathrm{sin}(\varphi),~\mathrm{cos}(\varphi),~0\Big] \\
    \boldsymbol{e}_{\mathrm{m}}^{\parallel} &= \Big[\mathrm{cos}(\theta_{\mathrm{m}})\mathrm{cos}(\varphi),~\mathrm{cos}(\theta_{\mathrm{m}})\mathrm{sin}(\varphi),~\mathrm{sin}(\theta_{\mathrm{m}})\Big] \\
    \boldsymbol{e}_{\mathrm{cl}}^{\parallel} &= \Big[-\mathrm{cos}(\theta_{\mathrm{cl}})\mathrm{cos}(\varphi),~-\mathrm{cos}(\theta_{\mathrm{cl}})\mathrm{sin}(\varphi),~-\mathrm{sin}(\theta_{\mathrm{cl}})\Big] \\
    \boldsymbol{s}_{\mathrm{m}} &= \Big[\mathrm{sin}(\theta_{\mathrm{m}})\mathrm{cos}(\varphi),~\mathrm{sin}(\theta_{\mathrm{m}})\mathrm{sin}(\varphi),~-\mathrm{cos}(\theta_{\mathrm{m}})\Big]~.
\end{align}
Here, $2~/~[1+\mathrm{cos}(\theta_{\mathrm{m}})]$ and $\sqrt{\mathrm{cos}(\theta_{\mathrm{cl}})}$ are the apodization factors of the parabolic mirror and collections lens, respectively. Defining the magnification factor as $M=f_{\mathrm{cl}}~/~f_{\mathrm{m}}$, we have
\begin{align}
    \theta_{\mathrm{m}} &= 2~\mathrm{arctan}\left(\frac{M~\mathrm{sin}(\theta_{\mathrm{cl}})}{2}\right) \\
    \theta_{\mathrm{cl}}^{\mathrm{max}} &= \mathrm{arctan}\left(\frac{2r_{\mathrm{m}}}{f_{\mathrm{cl}}}\right) ~ ,
\end{align}
where $M=25$ and $\theta_{\mathrm{cl}}^{\mathrm{max}} \approx 0.16$ radians for the particular experimental setup considered here. Taylor expanding the exponential term in Eq.~(\ref{FieldExpression}) to first-order about $\boldsymbol{q}=\boldsymbol{0}$, we find that we can write the scattered field as
\begin{align}\label{TaylorExpansion}
    \boldsymbol{E}_{\mathrm{scat}}(x, y~; \boldsymbol{q}(t)) &= \boldsymbol{E}_{\mathrm{scat}}(x, y ~;~ \boldsymbol{q} = \boldsymbol{0})~+ \stackrel{\leftrightarrow}{\hidewidth\boldsymbol{J}}_{\mathrm{scat}}\hspace{-1mm}(x, y ~;~ \boldsymbol{q} = \boldsymbol{0}) \boldsymbol{q}(t)^{T} ~ ,
\end{align}
where we have introduced the Jacobian tensor
\begin{align}
    \stackrel{\leftrightarrow}{\hidewidth\boldsymbol{J}}_{\mathrm{scat}}\hspace{-1mm}(x, y ~;~ \boldsymbol{q} = \boldsymbol{0}) &= \begin{bmatrix}
        \nabla^{T}\boldsymbol{E}_{\mathrm{scat}}(x, y ~;~ \boldsymbol{q} = \boldsymbol{0}) \cdot \hat{\boldsymbol{e}}_{x} \\
        \nabla^{T}\boldsymbol{E}_{\mathrm{scat}}(x, y ~;~ \boldsymbol{q} = \boldsymbol{0}) \cdot \hat{\boldsymbol{e}}_{y} \\
        \nabla^{T}\boldsymbol{E}_{\mathrm{scat}}(x, y ~;~ \boldsymbol{q} = \boldsymbol{0}) \cdot \hat{\boldsymbol{e}}_{z}
    \end{bmatrix} ~ ,
\end{align}
for 
\begin{align}
    \nabla^{T}f = \left[\frac{\partial f}{\partial(\boldsymbol{q}\cdot\hat{\boldsymbol{e}}_{x})}, \frac{\partial f}{\partial(\boldsymbol{q}\cdot\hat{\boldsymbol{e}}_{y})}, \frac{\partial f}{\partial(\boldsymbol{q}\cdot\hat{\boldsymbol{e}}_{z})}\right] ~ .
\end{align}
For the simplest case where the nanoparticle oscillates about a single spatial axis, such that $\boldsymbol{q}(t) = q(t)\hat{\boldsymbol{e}}_{q}$ for some principle axis $q = x, y$ or $z$, Eq.~(\ref{TaylorExpansion}) reduces to
\begin{align}
    \boldsymbol{E}_{\mathrm{scat}}(x, y~; \boldsymbol{q}(t)) &= \boldsymbol{E}_{\mathrm{scat}}(x, y~; \boldsymbol{0}) + q(t)\frac{\partial\boldsymbol{E}_{\mathrm{scat}}(x, y ~;~ \boldsymbol{0})}{\partial(\boldsymbol{q}\cdot\hat{\boldsymbol{e}}_{q})} \nonumber \\
    &= \boldsymbol{E}_{0}(x, y) + \frac{q(t)}{\lambda_{0}}\boldsymbol{E}^{(q)}(x, y) ~ , \label{ScatteredFieldDecomp}
\end{align}
where $\lambda_{0}$ refers to the free-space wavelength of the trapping beam, and has been introduced to normalize $q(t)$. Upon inspection, we find that this is equivalent to decomposing the total field into elastically and inelastically scattered contributions.

\section{Calculating the coupling efficiencies}\label{sec:CouplingCalcs}

\noindent To calculate the coupling of each scattered field component into the space-division demultiplexing photonic chip (spatial mode sorter), we need to formulate the guided modes. The input to the chip is an embedded few-mode, graded-index optical fiber (FMF) designed for operation at $\lambda = 1550$~nm, with a core radius ($r_{\mathrm{fib}}$) of approximately 4~\textmu m. Due to the low refractive index contrast between the refractive indices of the core ($n_{1}$) and cladding ($n_{2}$) we treat the guided modes as linearly-polarized (LP) modes, whose electric field distributions are fully described by the family of complex amplitude functions defined for $\sqrt{x^{2}+y^{2}} \geq r_{\mathrm{fib}}$:
\begin{equation}
    \mathrm{LP}_{\ell, m}(\boldsymbol{r}) =
    J_{\ell}\Big(\frac{u\sqrt{x^{2}+y^{2}}}{r_{\mathrm{fib}}}\Big)\mathrm{cos}\left(\ell\phi\right)~\mathrm{exp}\big[-i\beta_{\ell, m}z\big],
\end{equation}
and for $\sqrt{x^{2}+y^{2}} \geq r_{\mathrm{fib}}$:
\begin{equation}
    \mathrm{LP}_{\ell, m}(\boldsymbol{r}) =\frac{J_{\ell}(u)}{K_{\ell}(w)}K_{\ell}\Big(\frac{w\sqrt{x^{2}+y^{2}}}{r_{\mathrm{fib}}}\Big)\mathrm{cos}\left(\ell\phi\right)~\mathrm{exp}\big[-i\beta_{\ell,m}z\big],
\end{equation}
where we have defined the quantities 
\begin{align}
    u &= r_{\mathrm{fib}}\sqrt{n_{1}^{2}k^{2} - \beta_{\ell, m}^{2}} \\
    w &= r_{\mathrm{fib}}\sqrt{\beta_{\ell, m}^{2} - n_{2}^{2}k^{2}} ~ ,
\end{align}
in terms of the \textsl{propagation constant}, $\beta_{\ell, m}$. Here, $J_{\ell}(\bullet)$ and $K_{\ell}(\bullet)$ refer to Bessel functions of the first kind, and modified Bessel functions of the second kind, respectively. In our particular setup, the spatial profile of each LP mode is rotated anti-clockwise through an angle of approximately $\theta = 12^{\mathrm{o}}$ from the laboratory frame of reference. To account for this, we transform the vector field of each LP mode according to
\begin{align}
    \textbf{LP}_{\ell,m}(\boldsymbol{r}) &= R(\theta)~ \mathrm{LP}_{\ell,m}(\boldsymbol{r}') ~\hat{\boldsymbol{e}}_{x} ~ ,
\end{align}
where $\boldsymbol{r}'=R^{T}(\theta)~\boldsymbol{r}$ and we have assumed that we are concerned with only those LP modes polarised along the $\hat{\boldsymbol{e}}_{x}$ axis. To perform this calculation, we have made use of the usual rotation matrix
\begin{align}
    R(\theta) &= \Bigg[\begin{smallmatrix}
        \mathrm{cos}\left(\theta\right) & \mathrm{sin}\left(\theta\right) & 0 \\
        -\mathrm{sin}\left(\theta\right) & \mathrm{cos}\left(\theta\right) & 0 \\
        0 & 0 & 1
    \end{smallmatrix}\Bigg] ~ .
\end{align}
Additionally, to accommodate for the angled end-face of the FMF, the photonic chip is inclined by approximately $\phi =5^{\mathrm{o}}$ to the propagation axis. This introduces a weak longitudinally-polarised component to each LP mode. Hence, repeating the calculations above for a secondary rotation matrix
\begin{align}
    R(\phi) &= \Bigg[\begin{smallmatrix}
        1 & 0 & 0 \\
        0 & \mathrm{cos}\left(\phi\right) & \mathrm{sin}\left(\phi\right) \\
        0 & -\mathrm{sin}\left(\phi\right) & \mathrm{cos}\left(\phi\right)
    \end{smallmatrix}\Bigg] ~ ,
\end{align}
we may define a set of electric fields
\begin{align}\label{LPElectricFields}
    \textbf{LP}_{\ell, m}(\boldsymbol{r}) &=  \mathrm{LP}_{\ell, m}(\boldsymbol{r}'') \Bigg[\begin{smallmatrix}
        \mathrm{cos}\left(\theta\right) \\
        -\mathrm{sin}\left(\theta\right)\mathrm{cos}\left(\phi\right) \\
        \mathrm{sin}\left(\theta\right)\mathrm{sin}\left(\phi\right)
    \end{smallmatrix}\Bigg] ~ ,
\end{align}
in terms of the modified coordinates
\begin{align}
    \boldsymbol{r}'' &= \Bigg[\begin{smallmatrix}
        x~\mathrm{cos}\left(\theta\right) - y~\mathrm{sin}\left(\theta\right) \\
        \mathrm{cos}\left(\phi\right)[x~\mathrm{sin}\left(\theta\right) + y~\mathrm{cos}\left(\theta\right)] - z~\mathrm{sin}\left(\phi\right) \\
        \mathrm{sin}\left(\phi\right)[x~\mathrm{sin}\left(\theta\right) + y~\mathrm{cos}\left(\theta\right)] + z~\mathrm{cos}\left(\phi\right)
    \end{smallmatrix}\Bigg] ~ .
\end{align}
Importantly, these fields remain mutually orthogonal with respect to the normalized inner-product
\begin{align}\label{SpatialOverlapIntegral}
    \left|\iint_{-\infty}^{\infty} \textbf{LP}_{\ell_{1}, m_{1}}^{*}(x, y) \cdot \textbf{LP}_{\ell_{2}, m_{2}}(x, y) ~ \mathrm{d}x\mathrm{d}y\right| &= 0 ~ , 
\end{align}
for all $(\ell_{1}, m_{1}) \neq (\ell_{2}, m_{2})$. To calculate the coupling efficiency of the inelastically scattered field $\boldsymbol{E}^{(q)}(x, y)$ into any given LP mode, we calculate the quantity
\begin{align}\label{OverlapIntegral}
    \eta_{q}^{\ell,m} &= \frac{\left| \iint_{\mathrm{FMF}} \big[\boldsymbol{E}^{(q)}(x, y)\big]^{*} \cdot \textbf{LP}_{\ell, m}(x,y) ~ \mathrm{d}x\mathrm{d}y \right|^2}{\iint_{\mathrm{FMF}} \big|\boldsymbol{E}^{(q)}(x, y)\big|^{2}~\mathrm{d}x\mathrm{d}y \times \iint_{\mathrm{FMF}} \big|\textbf{LP}_{\ell, m}(x,y)\big|^{2}~\mathrm{d}x\mathrm{d}y} ~ ,
\end{align}
where integration is carried out over the input face of the few-mode input fiber (FMF). Finally, we note that Eq.~(\ref{OverlapIntegral}) is an approximate expression for the coupling coefficient, derived under the assumption that the coupled modes are exclusively transverse and spatially polarisation-matched~\cite{wagner_coupling_1982}, and that reflection from the face of the FMF is negligible. In this demonstration both conditions are approximately fulfilled, owing to the natural transverse characteristics of the LP modes, and the small contrast between air and fiber material ($n= 1.4679$). Exact analysis of the coupling, employing numerical modelling of the coupling between the scattered fields and an optimised set of fiber modes, will be published elsewhere.

\section{Deriving the measurement imprecision}\label{sec:MeasurementImp}

\noindent We must realistically consider the coupling of the total scattered field into the spatial mode sorter, not just the inelastically scattered term $\boldsymbol{E}^{(q)}(x, y)$. First, however, let us consider the form of the total electric field incident on the input plane of the spatial mode sorter. For the case where the nanoparticle oscillates only along the $\hat{\boldsymbol{e}}_{q}$ axis, this field is of the form
\begin{align}
    \boldsymbol{E}_{\mathrm{inc}}(x, y) &= \boldsymbol{E}_{\mathrm{ref}}(x, y) + \boldsymbol{E}_{0}(x, y) + \frac{q(t)}{\lambda_{0}}\boldsymbol{E}^{(q)}(x, y) ~ .
\end{align}
Moving forward, we choose to represent the electric field in terms of the complex normalised modes ${\hat{\boldsymbol{E}}(x, y) = \boldsymbol{E}(x, y) ~ / ~ (\iint_{\mathrm{FMF}} |\boldsymbol{E}(x, y)|^{2} ~ \mathrm{d}x\mathrm{d}y)^{1/2}}$ and corresponding complex amplitudes $\alpha$. To this end, we write
\begin{align}
    \boldsymbol{E}_{\mathrm{inc}}(x, y) &= \hat{\boldsymbol{E}}_{\mathrm{ref}}(x, y)\alpha_{\mathrm{ref}} + \hat{\boldsymbol{E}}_{0}(x, y)\alpha_{0} + \frac{q(t)}{\lambda_{0}}\hat{\boldsymbol{E}}^{(q)}(x, y)\alpha_{q} ~.
\end{align}
The electric field excited in the LP$_{\ell,m}$ channel of the spatial mode sorter is then
\begin{align}
\begin{split}
    \textbf{LP}_{\ell,m}(x, y) &= \hat{\textbf{LP}}_{\ell,m}(x, y) \left[ \sqrt{\eta^{\ell,m}_{\mathrm{ref}}}\alpha_{\mathrm{ref}} \right. \\
    & \left. + \sqrt{\eta^{\ell,m}_{0}}\alpha_{0} + \frac{q(t)}{\lambda_{0}}\sqrt{\eta^{\ell,m}_{q}}\alpha_{q} \right],
\end{split}
\end{align}
where $\eta^{\ell,m}_{\mathrm{ref}}$ and $\eta^{\ell,m}_{0}$ are defined analogously to Eq.~(\ref{OverlapIntegral}) except for the LO and elastically scattered fields, respectively. Neglecting any internal losses, we expect that the output power in the LP$_{\ell,m}$ channel measured by the detector can therefore be estimated as
\begin{equation}
\begin{split}
    P^{\ell,m}_{q}(t) &= \sqrt{\frac{\varepsilon_{0}}{\mu_{0}}}\left|\textbf{LP}_{\ell,m}(x, y)\right|^{2} \\
    & = \sqrt{\frac{\varepsilon_{0}}{\mu_{0}}} \Bigg(\eta^{\ell,m}_{\mathrm{ref}}|\alpha_{\mathrm{ref}}|^{2} + \eta^{\ell,m}_{0}|\alpha_{0}|^{2} + \eta^{\ell,m}_{q}\Big(\frac{q(t)}{\lambda_{0}}\Big)^2|\alpha_{q}|^{2} \\
    & +  \sqrt{\eta^{\ell,m}_{\mathrm{ref}}\eta^{\ell,m}_{0}}(\alpha_{\mathrm{ref}}\alpha_{0}^*+\alpha_{0}\alpha_{\mathrm{ref}}^*) \\
    & + \frac{q(t)}{\lambda_{0}}\sqrt{\eta^{\ell,m}_{q}}\Big(\alpha_{q}(\sqrt{\eta^{\ell,m}_{\mathrm{ref}}}\alpha_{\mathrm{ref}}^* \\
    & + \sqrt{\eta^{\ell,m}_{0}}\alpha_{0}^*) + \alpha_{q}^*(\sqrt{\eta^{\ell,m}_{\mathrm{ref}}}\alpha_{\mathrm{ref}} + \sqrt{\eta^{\ell,m}_{0}}\alpha_{0})\Big)\Bigg),\label{RecoveredSignal}
\end{split}
\end{equation}

In the following, we discard the term proportional to $(q(t)/\lambda_0)^{2}$ as we expect the displacements to be far smaller than $\lambda_0$ along each axis. Note that, from Eq.~(\ref{RecoveredSignal}), we do not simply measure $q(t)$. Instead, the measurement result is dominated by a series of time-independent coefficients, which mask the information available on the nanoparticle's position. From the form of these coefficients, we can estimate the measurement imprecision.\\
All terms contributes to shot noise and noting that $\alpha_{\mathrm{ref}},\alpha_{0}\gg(q(t)/\lambda_0)\alpha_{q}$ we can write the power spectral density (PSD) of the noise as:

\begin{align}\label{ShotNoisePSD}
\begin{split}
    S^{\ell, m}_{\mathrm{noise}} = \frac{\hbar k c}{2\pi}\sqrt{\frac{\varepsilon_{0}}{\mu_{0}}}&\Bigg(\eta^{\ell, m}_{\mathrm{ref}} |\alpha_{\mathrm{ref}}|^{2} + \eta^{\ell,m}_{0} |\alpha_{0}|^{2}  \\
    &+ \sqrt{\eta^{\ell,m}_{\mathrm{ref}}\eta^{\ell,m}_{0}}(\alpha_{\mathrm{ref}}\alpha_{0}^* + \alpha_{0}\alpha_{\mathrm{ref}}^*) \Bigg)
\end{split}
\end{align}
Additionally, to extract imformation on the position $q(t)$ we need to weight it by a factor of
\begin{align}\label{BetaDef}
\begin{split}
    \beta_{q}^{\ell,m} = \frac{1}{\lambda_{0}}\sqrt{\frac{\varepsilon_{0}}{\mu_{0}}}\sqrt{\eta^{\ell,m}_{q}}&\Bigg(\alpha_{q}\Big(\sqrt{\eta^{\ell,m}_{\mathrm{ref}}}\alpha_{\mathrm{ref}}^* \\
    &+ \sqrt{\eta^{\ell,m}_{0}}\alpha_{0}^*\Big) + \alpha_{q}^*\Big(\sqrt{\eta^{\ell,m}_{\mathrm{ref}}}\alpha_{\mathrm{ref}} + \sqrt{\eta^{\ell,m}_{0}}\alpha_{0}\Big) \Bigg)
\end{split}
\end{align}
which is squared when calculating the PSD. The imprecision associated with measuring the nanoparticle's motion along the $\boldsymbol{\hat{e}}_{q}$ axis, in the LP$_{\ell,m}$ output channel of the spatial mode sorter, will therefore be of the form ${S_{\mathrm{imp}, q}^{\ell,m} = S^{\ell,m}_{\mathrm{noise}} ~ / ~ (\beta_{q}^{\ell,m})^{2}}$~\cite{tebbenjohanns_optimal_2019, maurer_quantum_2023}. \\
Let's examine closer the form of the imprecision noise in the case where the LO is effectively decoupled from the LP mode ($\eta^{\ell,m}_{\mathrm{ref}} = 0$). In this case, the measurement imprecision reduces to
\vspace{1.5mm}
\begin{align}\label{MeasurementImprecisionForChip}
    S^{\ell, m}_{\mathrm{imp}, q} &= \frac{\hbar k c}{8\pi}\sqrt{\frac{\mu_{0}}{\varepsilon_{0}}} \frac{1}{|\alpha_{q}|^{2}}\frac{\lambda_{0}^{2}}{\eta^{\ell,m}_{q}} ~ ,
\end{align}
which is minimised when $\eta^{\ell,m}_{q} = 1$. Perhaps counter-intuitively, this result suggests that the measurement imprecision is independent of the overlap of the inelastically and elastically scattered field, despite Eq.~(\ref{RecoveredSignal}) suggesting that our measurement relies on the interference of these two fields. \\
Remarkably, in the case where we have a strong LO that is coupled to the LP channels ($\alpha_{\mathrm{ref}} \gg \alpha_{0}, \alpha_{q} $), we arrive at the same result where the measurement imprecision is independent of the overlap of the inelastically scattered fields and the LO.

\begin{figure*}[ht]
    \centering
        \includegraphics[width=\linewidth]{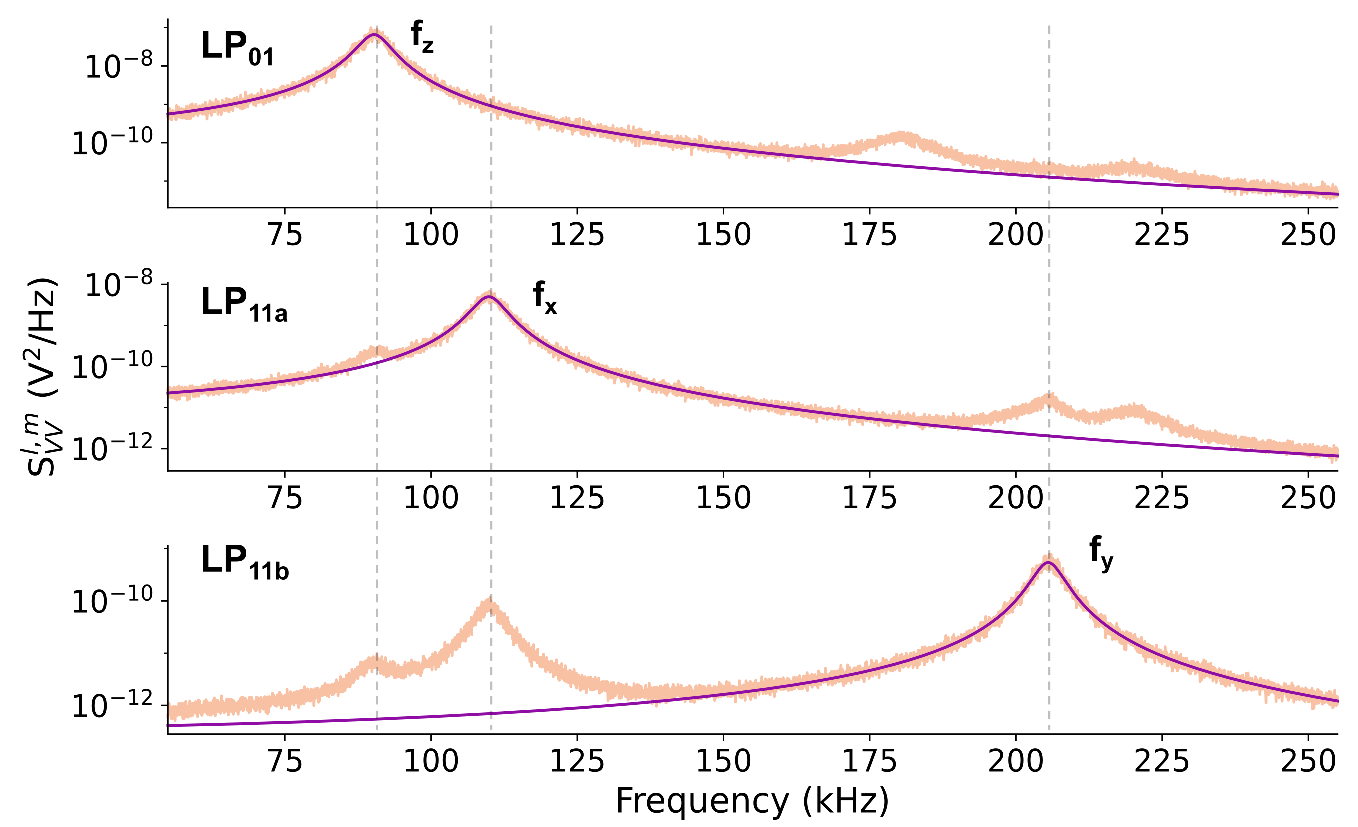}
        \caption{
        \textbf{(a)} Raw power spectral densities (PSDs) at the LP$_{01}$, LP$_{11\mathrm{a}}$ and LP$_{11\mathrm{b}}$ output channels of the spatial mode sorter at a pressure of 11.3 mbar. The fits are calculated using Eq.~(\ref{MotionalPSD}).
        }
        \label{FigCalib}
\end{figure*}

\section{Form of the power spectral density and calculation of calibration factor}\label{sec:FormOfPSD}

\noindent Let us now consider the form of the power spectral density (PSD) measured at each LP$_{\ell, m}$ output channel of the space-division de-multiplexing photonic chip. In Appendix~\ref{sec:MeasurementImp} we argued that the power measured at a detector is given by Eq.~(\ref{RecoveredSignal}), and can be split into those terms independent of the nanoparticle's motion, and those terms proportional to $q(t)$. We have already identified the former as sources of noise in the measurement result, with a PSD given by Eq.~(\ref{ShotNoisePSD}). Collecting the remaining terms, we introduce $s_{qq}^{\ell,m}(t) = q(t) \beta_{q}^{\ell,m}$, where $\beta_{q}^{\ell,m}$ is defined in Eq.~(\ref{BetaDef}). From the Wiener-Khinchin theorem, this quantity has a PSD of the form
\begin{align}
    S_{qq}^{\ell,m}(\Omega) &= \int^{\infty}_{-\infty} \langle s_{qq}^{\ell,m}(t)s_{qq}^{\ell,m}(t+\tau)\rangle e^{-i\Omega\tau}~\mathrm{d}\tau \nonumber \\
    &= (\beta_{q}^{\ell,m})^{2}~\int^{\infty}_{-\infty} \langle q(t)q(t+\tau)\rangle e^{-i\Omega\tau} ~ \mathrm{d}\tau ~ .
\end{align}
The second term in this expression reduces to the well-known result
\begin{align}
    S_{qq}^{\ell,m}(\Omega) &= (\beta_{q}^{\ell,m})^{2} \frac{k_{B}T_{0}}{\pi m}\frac{\gamma}{\left(\Omega_{q}^{2}-\Omega^{2}\right)^{2} + \Omega^{2}\gamma^{2}} ~ .
\end{align}
The total PSD measured at each LP output channel of the space-division de-multiplexing photonic chip is then of the form
\begin{align}
    S^{\ell,m}_{\mathrm{tot}}(\Omega) &= S^{\ell,m}_{\mathrm{imp}, q} + S^{\ell,m}_{qq}(\Omega) ~ .
\end{align}
Here we see the role of the imprecision in this measurement, derived in Appendix~(\ref{sec:MeasurementImp}), which functions as the noise floor of the PSD.\\
\noindent In order to calibrate our measurement, we follow the method oulined in \cite{hebestreit_calibration_2018}. We record PSDs, showed in Fig.~\ref{FigCalib},  at pressure where the nanomechanical oscillator is thermalised with the environment through gas collision (P $>$ 10 mbar). The PSDs used for calibration are taken after three cycles consisting of pumping down to 10$^{-3}$ mbar and subsequently filling the chamber with N$_2$ up to 500 mbar. \\
We fit the PSD of each channel with the following Lorentzian function derived from the equation of motion of a thermally damped oscillator:
\begin{align}\label{MotionalPSD}
    S_{VV}(\Omega) &= c_{calib}^{2}\frac{k_{B}T_{0}}{\pi m}\frac{\gamma}{\left(\Omega_{q}^{2}-\Omega^{2}\right)^{2} + \left(\gamma\Omega\right)^{2}} ~ ,
\end{align}
where $\gamma$ is the damping rate and $\Omega_{q}$ the oscillation frequency of the $\it{q}$ axis. The relation between the uncalibrated and calibrated PSDs is simply given by $S_{VV}(\Omega) = c_{calib}^{2}S_{qq}(\Omega)$.
At pressure where the main contribution to damping comes from gas collisions, we can infer the size of the nanoparticle from the damping rate using kinetic theory and knowing the density of silica.\\
Finally in order to avoid error in the calibration due to anharmonicity of the potential, we calculate the PSD of the velocity from the PSD of the position with $S_{\dot{q}\dot{q}}(\Omega) = \Omega^{2}S_{qq}(\Omega)$.
From the Wiener-Khinchin theorem, we can calculate the variance of the velocity (or position) by integrating the PSD around $\Omega_{q}$ (bandwith of 20 kHz). Then using the equipartition theorem for the kinetic energy we can then extract the calibration factor:
\begin{align}\label{equip_theo_kin}
    \frac{1}{2}m\langle\dot{q}\rangle=\frac{1}{2}m\frac{\langle\dot{V}\rangle}{c_{calib}^2}=\frac{1}{2}k_BT_{0}
\end{align}
\\
We also use Eq.\ref{equip_theo_kin} to estimate the temperature under feedback cooling with the calibration factors calculated at room temperature for each DOFs in their respective detection channel.

\section{Discussion of losses and efficiencies}\label{sec:DiscussionOfLoss}

Here, we provide a brief overview of the efficiencies reported in the main text.

\begin{table}[ht]
    \centering
    \begin{tabular}{c c}
        \hline
        \hline
        \textbf{Information Efficiencies} & \textbf{Values} \\
        \hline 
        \hline \noalign{\smallskip}
        Collection Efficiency & $\begin{cases}
            0.50 & ,~ x(t) \\
            0.50 & ,~ y(t) \\
            0.90 & ,~ z(t) \\
        \end{cases}$ \\
        Path Losses & $0.90$ \\
        Insertion Losses & $0.79$ \\
        Polarization-dependent Losses & $0.99$ \\
        Coupling Efficiency & $\begin{cases}
            0.61 & ,~ x(t) \to \mathrm{LP}_{11\mathrm{a}} \\
            0.82 & ,~ y(t) \to \mathrm{LP}_{11\mathrm{b}} \\
            0.86 & ,~ z(t) \to \mathrm{LP}_{01} \\
        \end{cases}$ \\
        Quantum Efficiency & 0.80 \\
        Dark Noise & 0.79 \\
        Digital Noise & 0.98 \\
        \hline \noalign{\smallskip}
        Measurement Efficiency $\left(\eta^{(q)}_{\mathrm{tot}}\right)$ & $\begin{cases}
            0.13 & ,~ x(t) \to \mathrm{LP}_{11\mathrm{a}} \\
            0.18 & ,~ y(t) \to \mathrm{LP}_{11\mathrm{b}} \\
            0.33 & ,~ z(t) \to \mathrm{LP}_{01}
        \end{cases}$ \\
        \noalign{\smallskip} \hline
        \hline
    \end{tabular}
    \caption{
        Total efficiencies for the setup considered here.
    }
    \label{tab:LossesTableMainText}
\end{table}
\subsubsection{Collection efficiency}
\noindent We write the effective numerical aperture (NA) of the parabolic mirror as 
\begin{align}
    \mathrm{NA} &= 1 - \mathrm{cos}\left(\mathrm{arctan}\left[\frac{r_{\mathrm{m}}}{f_{\mathrm{m}} - \frac{r_{\mathrm{m}}^{2}}{4f_{\mathrm{m}}}}\right]\right) ~ , 
\end{align}
where $f_{\mathrm{m}}$ and $r_{\mathrm{m}}$ are the focal length (1~mm) and radius (2~mm) of the parabolic mirror, respectively. For the design specifications used here, this implies a numerical aperture of $\mathrm{NA}= 1$. Hence, we expect to collect all of the light back-scattered from an optically-levitated nanoparticle, equivalent to a photon collection efficiency of $50\%$ for all three center-of-mass (COM) translational degrees of freedom (DOFs). In the context of information losses, however, we require a more nuanced approach. This is because motional information is not uniformly distributed in the scattered field. To this end, we consider the so-called \textsl{information radiation patterns} (IRPs) associated with each COM translational DOF, $q(t)$ for $q = x,y,z$~\cite{tebbenjohanns_optimal_2019, maurer_quantum_2023}. These IRPs are of the form
\begin{align}
    \mathcal{I}_{x}(\theta, \varphi) &= \frac{15}{8\pi}\Big[1-\mathrm{sin}^{2}(\theta)\mathrm{cos}^{2}(\phi)\Big]~\mathrm{sin}^{2}(\theta)\mathrm{cos}^{2}(\varphi)~\mathrm{d}\Omega \\
    \mathcal{I}_{y}(\theta, \varphi) &= \frac{15}{16\pi}\Big[1-\mathrm{sin}^{2}(\theta)\mathrm{cos}^{2}(\phi)\Big]~\mathrm{sin}^{2}(\theta)\mathrm{sin}^{2}(\varphi)~\mathrm{d}\Omega \\
    \mathcal{I}_{z}(\theta, \varphi) &= \frac{3}{8\pi(\frac{2}{5}+A^{2})}\Big[1-\mathrm{sin}^{2}(\theta)\mathrm{cos}^{2}(\phi)\Big]\Big(\mathrm{cos}(\theta)-A\Big)^{2}~\mathrm{d}\Omega ~ ,
\end{align}
for an $x$-polarised dipolar scatterer. Here, $A$ is a consequence of the Gouy phase inherited from the trapping beam. Note that, for intermediate values of the numerical aperture, we find $A \approx 1-(kz_{\mathrm{R}})^{-1}$ for Rayleigh length $z_{\mathrm{R}}$. Given that our setup operates well beyond this limit, however, we choose to focus on illumination by a plane wave ($A=1$). To determine the information collection efficiency for the parabolic mirror, we evaluate the integral
\begin{align}
    \eta_{q} &= \int^{2\pi}_{0}\int^{\pi}_{\pi/2}  \mathcal{I}_{q}(\theta, \varphi)~\mathrm{d}\theta\mathrm{d}\varphi ~ ,
\end{align}
where the lower integration limit on $\theta$ is given by $\pi - \mathrm{arcsin}(\mathrm{NA})$. This represents the amount of information content collected in the back focal plane of the optical trap, and gives $(\eta_{x}, \eta_{y}, \eta_{z}) = (50, 50, 90)\%$, respectively.
\subsubsection{Path losses}
\noindent To estimate the path losses, we measure the difference in power of the forward- and backward-propagating trapping beam in the focal plane of the collection lens, with no nanoparticle trapped. To this end, we measure the losses at each optical interface from the collimation lens to the parabolic mirror. From this, we arrive at a value for the path losses of approximately $10\%$, equivalent to an efficiency of $90\%$.

\begin{figure*}[ht]
    \centering
        \includegraphics[width=\linewidth]{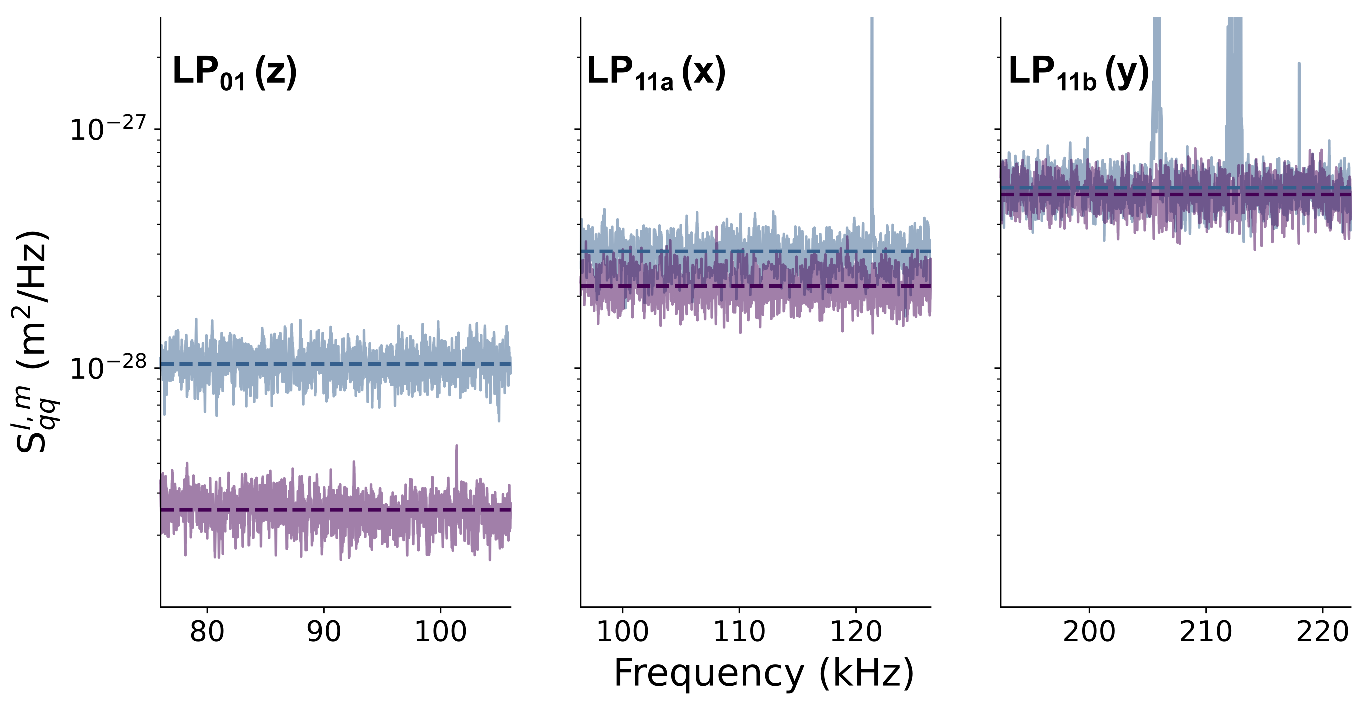}
        \caption{
        PSDs of the empty trap (grey) and detector noise (purple). The average of each curve is showed as a dashed line.
        }
        \label{noise_zoom}
\end{figure*}

\subsubsection{Coupling efficiency}
\noindent Inspired by the ability of our numerical model to predict the selectivity of the spatial mode sorter, we use those coupling efficiencies reported in the main text.
\subsubsection{Quantum efficiency}
\noindent From the technical specifications provided by the supplier of the detectors used here (\textsl{Thorlabs, PDB450C}), we estimate the quantum efficiency of our detectors to be $0.80\%$ at $\lambda = 1550$~nm.
\subsubsection{Dark noise}
\noindent Information loss must also be considered in the electronic line, after each signal has been measured. In particular, we note that the dark current noise will mask some of the available information. At frequencies near to the natural oscillation frequencies of the nanoparticle, we measure the dark noise to be approximately $8$~dB below the shot noise, equivalent to an efficiency of $79\%$. 
\subsubsection{Digital noise}
\noindent In the context of measurement-based control techniques, it is meaningful to consider those additional losses which stem from use of a digital feedback loop. In particular, here we are interested in using a Field Programmable Gate Array (FPGA) board (\textsl{STEMlab 125-14, Red Pitaya}) which functions as a phase-locked loop (PLL), with an estimated efficiency of $95\%$.

\section{Comparison of Measured and Predicted Efficiencies}\label{sec:EfficiencyComparison}

From the calibrated PSDs we can directly measure the measurement imprecision, which is equal to the noise floor shown in Fig.~2, 3 and Fig.~\ref{noise_zoom} in the limit of low pressures ($\leq 10^{-8}$~mbar), where contributions from random collisions with background gas molecules are negligible and the total noise floor of the system converges to the shot noise. In particular, we are interested in the square-root of the imprecision, which determines the position resolution of our detection system. For the data presented here, if we take the empty trap as the imprecision noise, we measure  $(S_{\mathrm{imp}, x}^{11\mathrm{a}}, S_{\mathrm{imp}, y}^{11\mathrm{b}}, S_{\mathrm{imp}, z}^{01}) = (3.1, 5.7, 1.0)\pm(0.8, 1.4, 0.5) \times 10^{-28}$~m$^{2}$~/~Hz. This implies we can resolve displacements of $(17.5, 23.8, 10.0)\pm(0.9, 1.3, 0.5)$~fm~/~$\sqrt{\mathrm{Hz}}$ along the $x$, $y$ and $z$-axes, respectively. Importantly, these values are below the predicted zero-point motion of the nanoparticle, $(x_{\mathrm{zpf}}, y_{\mathrm{zpf}}, z_{\mathrm{zpf}}) = (2.0, 1.4, 2.2)$~pm.

To independently confirm the measurement efficiencies reported in Table~S1 are actually representative of the setup reported on here, we compare the predicted position resolution to the value obtained above. That is, we expect the measurement imprecision to go as~\cite{dania_position_2022}
\begin{align}\label{efficiencycheck}
    S_{\mathrm{imp},q}^{\ell,m} &= \frac{5\hbar c\lambda_{0}}{8\pi\eta^{\ell,m}_{\mathrm{tot},q}P_{\mathrm{scat}}} ~ ,
\end{align}
where $P_{\mathrm{scat}}$ is the optical power scattered from a trapped nanoparticle. In the Rayleigh approximation, this can be calculated as $P_{\mathrm{scat}} = I_{0}\sigma$, where $I_{0} = 2P_{0}~/~\pi w_{0}^{2}$ can be written in terms of the power ($P_{0}$) and waist ($w_{0}$) of the trapping beam at the focus of the parabolic mirror. Further,
\begin{align}
    \sigma &= \frac{8\pi}{3}\Bigg(\frac{\alpha_{\mathrm{p}}k^{2}}{4\pi\varepsilon_{0}}\Bigg)^{2} \\
    \alpha_{\mathrm{p}} &= 4\pi r_{\mathrm{p}}^{3}\varepsilon_{0}\frac{n_{\mathrm{p}}^{2}-1}{n_{\mathrm{p}}^{2}+2} ~ ,
\end{align}
are the nanoparticle's scattering cross-section and dipolar polarisability, respectively. Importantly, the relationship presented in Eq.~(\ref{efficiencycheck}) is agnostic of the particular experimental setup used. For the efficiencies reported in Table 1 we find predicted measurement imprecisions of $(S_{\mathrm{imp}, x}^{11\mathrm{a}}, S_{\mathrm{imp}, y}^{11\mathrm{b}}, S_{\mathrm{imp}, z}^{01}) = (2.3, 2.0, 1.0) \times 10^{-28}$~m$^{2}$~/~Hz, equivalent to position resolutions of $(15, 14.1, 9.9)$~fm~/~$\sqrt{\mathrm{Hz}}$ for the $x$, $y$ and $z$-axes, respectively. The difference between the predicted and measured values, particularly for the transverse DOFs (x and y) can be accounted for by a small angular mismatch of the DMX LP11 axis with respect to the levitated dipole, as well as imperfect coupling due to optical misalignment.

\bibliography{Paper}

\end{document}